\pdfoutput=1

\documentclass[11pt,twoside,a4paper,cmspaper,final,collab]{cms-tdr}
\def\svnVersion{369630}\def\cmsCernNoTag{CERN-EP/2016-013}\def\cmsCernDate{\today}\def\cmsMessage{Published in the European Physical Journal C as \href{http://dx.doi.org/10.1140/epjc/s10052-016-4346-8}{\doi{10.1140/epjc/s10052-016-4346-8.}}}

\begin{document}\cmsNoteHeader{SMP-14-015}

\hyphenation{had-ron-i-za-tion}
\hyphenation{cal-or-i-me-ter}
\hyphenation{de-vices}
\RCS$Revision: 348797 $
\RCS$HeadURL: svn+ssh://svn.cern.ch/reps/tdr2/papers/SMP-14-015/trunk/SMP-14-015.tex $
\RCS$Id: SMP-14-015.tex 348797 2016-06-22 21:11:31Z panos $

\newlength\cmsFigWidth
\ifthenelse{\boolean{cms@external}}{\setlength\cmsFigWidth{0.98\columnwidth}}{\setlength\cmsFigWidth{0.7\textwidth}}
\ifthenelse{\boolean{cms@external}}{\providecommand{\cmsLeft}{top\xspace}}{\providecommand{\cmsLeft}{left\xspace}}
\ifthenelse{\boolean{cms@external}}{\providecommand{\cmsRight}{bottom\xspace}}{\providecommand{\cmsRight}{right\xspace}}

\providecommand{\alps}{\ensuremath{\alpha_S}\xspace}
\providecommand{\alpsmz}{\ensuremath{\alpha_S(M_{\Z})}\xspace}
\providecommand{\chisq}{\ensuremath{\chi^2}\xspace}
\providecommand{\chisqndof}{\ensuremath{\chi^2/n_\mathrm{dof}}\xspace}
\providecommand{\NLOJETPP} {{\textsc{NLOJet++}}\xspace}
\providecommand{\fastNLO} {{\textsc{fastNLO}}\xspace}
\providecommand{\PYTHIAS} {{\textsc{pythia6}}\xspace}
\providecommand{\PYTHIAE} {{\textsc{pythia8}}\xspace}
\providecommand{\RooUnfold} {{\textsc{RooUnfold}}\xspace}
\providecommand{\MGaMCNLO} {{\textsc{MadGraph5\_aMC@NLO}}\xspace}
\providecommand{\NJet} {{\textsc{NJet}}\xspace}
\providecommand{\ptmax}{\ensuremath{\pt^{\text{max}}}\xspace}
\providecommand{\mur}{\ensuremath{\mu_r}\xspace}
\providecommand{\muf}{\ensuremath{\mu_f}\xspace}
\providecommand{\rbtrr}{\rule[-0.8ex]{0ex}{3.2ex}}
\providecommand{\dphi}{\ensuremath{\Delta\phi_\text{dijet}}\xspace}
\providecommand{\met}{\ETslash\xspace}
\providecommand{\metoset}{\ensuremath{\ETslash /\sum{\ET}  }\xspace}

\cmsNoteHeader{SMP-14-015}
\title{Measurement of dijet azimuthal decorrelation in pp~collisions at $\sqrt{s}=8\TeV$}

\date{\today}

\abstract{A measurement of the decorrelation of azimuthal angles between
  the two jets with the largest transverse momenta is presented for seven
  regions of leading jet transverse momentum up to 2.2\TeV. The analysis
  is based on the proton-proton collision data collected with the
  CMS experiment at a centre-of-mass energy of 8\TeV corresponding
  to an integrated luminosity of 19.7\fbinv.
  The dijet azimuthal decorrelation is caused by
  the radiation of additional jets and probes the dynamics of multijet
  production. The results are compared to fixed-order predictions of
  perturbative quantum chromodynamics (QCD), and to simulations using
  Monte Carlo event generators that include parton showers,
  hadronization, and multiparton interactions. Event generators with
  only two outgoing high transverse momentum partons fail to describe
  the measurement, even when supplemented with next-to-leading-order
  QCD corrections and parton showers. Much better agreement is
  achieved when at least three outgoing partons are complemented
  through either next-to-leading-order predictions or parton
  showers. This observation emphasizes the need to improve
  predictions for multijet production.}

\hypersetup{%
  pdfauthor={CMS Collaboration},%
  pdftitle={Measurement of dijet azimuthal decorrelation in pp collisions at sqrt(s) = 8 TeV},%
  pdfsubject={CMS},%
  pdfkeywords={CMS, physics, QCD, jets, dijets, azimuthal decorrelation}
}

\maketitle

\section{Introduction}
\label{sec:intro}

Hadronic jets with large transverse momenta \pt are produced in
high-energy proton-proton collisions when two partons interact
with high momentum transfer via the strong force.
At leading order (LO) in perturbative quantum chromodynamics (pQCD),
two final-state partons are produced back-to-back in the transverse plane.
For this case, the azimuthal angular
separation between the two leading \pt jets in the transverse plane,
$\dphi=\abs{\phi_\mathrm{jet1}-\phi_\mathrm{jet2}}$, equals~$\pi$.
The nonperturbative effects of multiparton
interactions or hadronization disturb this correlation only
mildly, and $\dphi\approx\pi$ still holds. However, the production of a third
high-\pt jet leads to a decorrelation in azimuthal
angle. The smallest achievable value of $\dphi=2\pi/3$ occurs
in a symmetric star-shaped 3-jet configuration. Fixed-order
calculations in pQCD for 3-jet production with up
to four outgoing partons provide next-to-leading-order
(NLO) predictions for the region of $2\pi/3 \leq \dphi < \pi$. If more than three jets are
produced, the azimuthal angle between the two leading jets can approach
zero, although very small angular separations are suppressed because
of the finite jet sizes for a particular jet algorithm.
The measurement of the dijet azimuthal angular decorrelation
is an interesting tool to gain insight into multijet
production processes without measuring jets beyond the
leading two.

This paper reports the measurement of the normalized dijet differential cross section as
a function of the dijet azimuthal angular separation,
\begin{equation}
  \frac{1}{\sigma_\text{dijet}}
  \frac{\rd\sigma_\text{dijet}}{\rd\dphi},
  \label{eq:dijet}
\end{equation}
for seven regions of the leading jet \pt, \ptmax, within a rapidity region
of $\abs{y}<2.5$.
Experimental and theoretical uncertainties are reduced
by normalizing the \dphi distribution to the total dijet cross section
$\sigma_\text{dijet}$ within each region of \ptmax.
For the first time, azimuthal angular separations \dphi over the full
phase space from 0 to $\pi$ are covered.
Comparisons are made to fixed-order predictions up to NLO for 3-jet production,
and to NLO and LO dijet as well as to tree-level multijet production,
each matched with parton showers and complemented with multiparton interactions
and hadronization.

The measurement is performed using data collected during 2012 with the
CMS experiment at the CERN LHC, corresponding
to an integrated luminosity of 19.7\fbinv of proton-proton
collisions at $\sqrt{s}=8\TeV$.
Previous measurements of dijet azimuthal decorrelation were
reported by the D0 Collaboration in $\Pp\PAp$
collisions at $\sqrt{s}=1.96\TeV$ at the
Tevatron \cite{bib:D0_1,bib:D0_2}, and by the CMS and ATLAS
Collaborations in $\Pp\Pp$ collisions at
$\sqrt{s}=7\TeV$ at the LHC \cite{bib:CMS,bib:ATLAS}.

\section{The CMS Detector}
\label{sec:detector}

A detailed description of the CMS detector, together with a definition of the coordinate system 
used and the relevant kinematic variables, can be found in Ref.~\cite{Chatrchyan:2008aa}.
The central feature of the CMS detector is a superconducting
solenoid, 13\unit{m} in length and 6\unit{m} in inner diameter, providing
an axial magnetic field of 3.8\unit{T}.
Within the field volume are a silicon pixel and strip tracker,
a lead tungstate crystal electromagnetic calorimeter (ECAL)
and a brass and scintillator hadron calorimeter (HCAL), each composed
of a barrel and two endcap sections.
Charged particle trajectories are measured by the tracker
with full azimuthal coverage within pseudorapidities $\abs{\eta}< 2.5$.
The ECAL, which is equipped with a preshower detector in the endcaps,
and the HCAL cover the region $\abs{\eta}< 3$.
In addition to the barrel and endcap detectors, CMS has extensive
forward calorimetry, which extends the coverage up to $\abs{\eta}< 5$.
Finally, muons are measured up to $\abs{\eta}< 2.4$ by gas-ionization
detectors embedded in the steel flux-return yoke outside the solenoid.

\section{Event reconstruction and selection}
\label{sec:selectionReconstruction}

This measurement uses data samples that were collected with single-jet high-level triggers
(HLT) \cite{CMS_HLT}. Four such single-jet HLTs were considered that
require at least one jet in the event to have $\pt > 140$, 200, 260, and $320 \GeV$, respectively.
All triggers were prescaled during the 2012 run except the highest-threshold trigger.
The integrated luminosity \lumi for the four trigger samples is shown in
Table~\ref{tbl:TrigLumi}.
The trigger efficiency is estimated using triggers with lower \pt thresholds.
Using these four jet-energy thresholds gives 100\% trigger efficiencies in the corresponding
four momentum regions $200<\ptmax<300\GeV$, $300<\ptmax<400\GeV$,
$400<\ptmax<500\GeV$, and $\ptmax>500\GeV$.

\begin{table}[htbH]
  \centering
    \topcaption{The integrated luminosity for each trigger
      sample considered in this analysis.\label{tbl:TrigLumi}}
    \begin{tabular}{lccccc}
      \hline
      HLT \pt threshold (\GeVns{})   &  140   &  200   &  260  & 320   \\
      \hline
      \lumi ($\text{fb}^{-1}$)\rbtrr &  0.06  &  0.26  &  1.06 & 19.7  \\
      \hline
    \end{tabular}

\end{table}

Particles are reconstructed and identified using a particle-flow
(PF) algorithm, which combines the information from the individual
subdetectors \cite{bib:pf1,bib:pf2}.
The four-vectors of particle candidates, reconstructed by the above technique,
are used as input to the jet-clustering algorithm.
Jets are reconstructed using the infrared- and collinear-safe
anti-\kt
clustering algorithm with a distance parameter $R=0.7$ \cite{bib:antikt}.
The clustering is performed with the \FASTJET package \cite{bib:fastjet}
using four-momentum summation.

The reconstructed jets require small additional energy corrections to account for various
reconstruction inefficiencies in tracks and clusters in the PF algorithm.
These jet energy corrections \cite{bib:cms-jes} are derived using (1) simulated events,
generated with \PYTHIA 6.4.22 \cite{bib:pythia6} with tune Z2* \cite{bib:Z2starCMS,bib:GEN-14-001}
and processed through the CMS detector simulation based on \GEANTfour \cite{bib:geant},
and (2) measurements containing dijet, photon+jet, and Z+jet events.
The jet energy corrections, which depend on the $\eta$ and \pt of the jet, are applied
to the jet four-momentum vectors as multiplicative factors \cite{bib:JES_2012}.
The overall factor is typically 1.2 or smaller, approximately
uniform in $\eta$, and is 1.05 or smaller for jets having $\pt > 100\GeV$.
An offset correction is applied to take into account the extra energy clustered
into jets from additional proton-proton interactions within the same or neighbouring bunch
crossings (in-time and out-of-time pileup) \cite{bib:cms-jes}.
Pileup effects are important only for jets with low \pt
and become negligible for jets with $\pt > 200 \GeV$.
The current measurement is, therefore, insensitive to pileup effects
on jet energy calibration.

Each event is required to have at least one vertex reconstructed offline \cite{bib:vertex_paper} with
a position along the beam line that is within 24\unit{cm} of the nominal interaction point.
To suppress nonphysical jets, \ie jets resulting from noise in the ECAL and/or HCAL calorimeters,
stringent criteria \cite{bib:jetID_PF} are applied for identifying jets: each jet should contain at least two particles,
one of which is a charged hadron, and the jet energy fraction carried by neutral hadrons
and photons should be less than 90\%.
The efficiency for identifying physical jets using these criteria is greater than~99\%.

The two leading jets, which define \dphi, are selected by considering
all jets in the event with $\pt > 100 \GeV$ and an absolute rapidity
$\abs{y}<5$.
Events are selected in which the leading jet \pt exceeds 200\GeV
and the rapidities $y_1$ and $y_2$ of the two leading jets lie within the tracker coverage
of $\abs{y}<2.5$.

\begin{figure}[hbtp]
  \centering
  \includegraphics[width=0.48\textwidth]{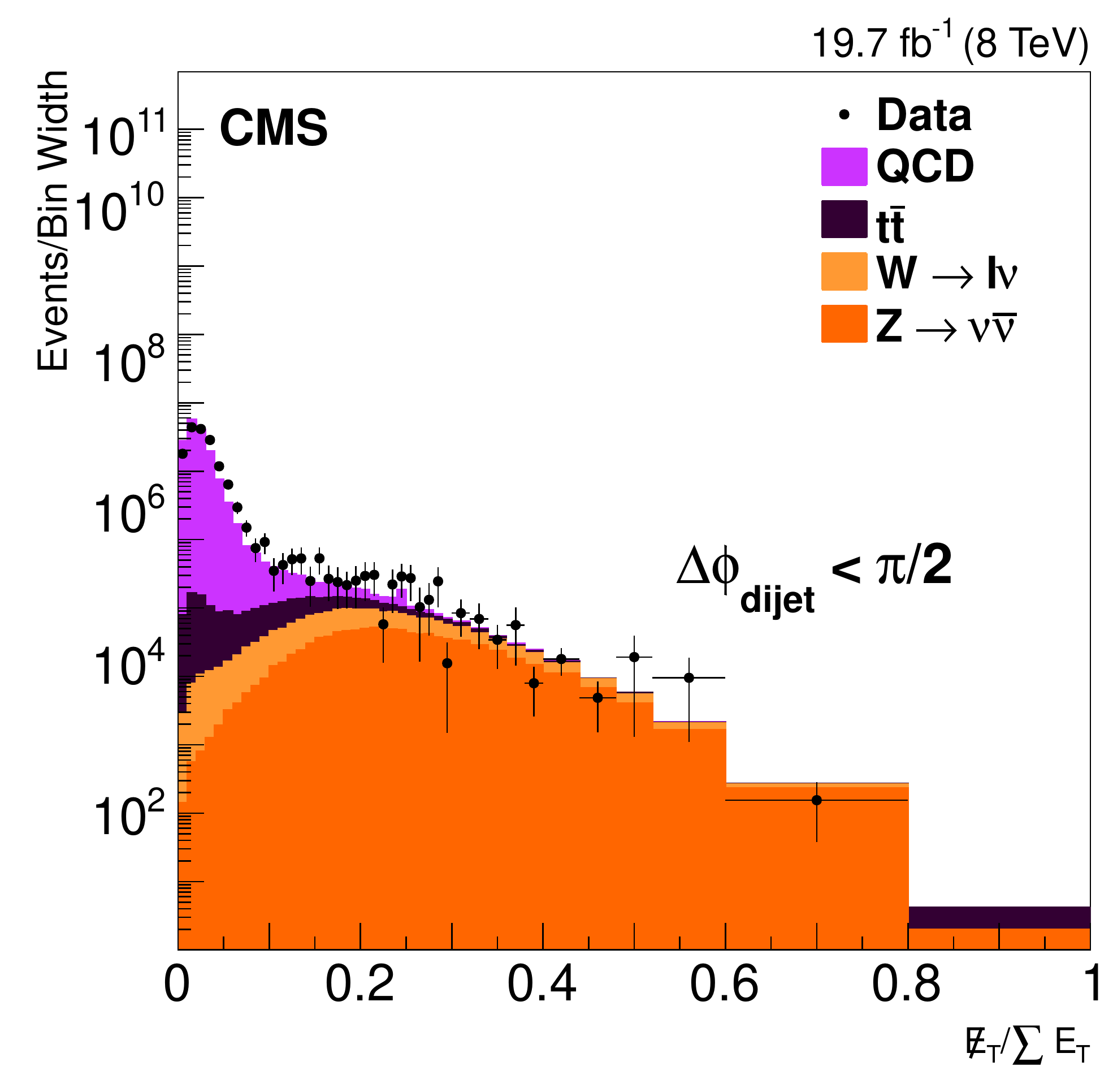}
  \includegraphics[width=0.48\textwidth]{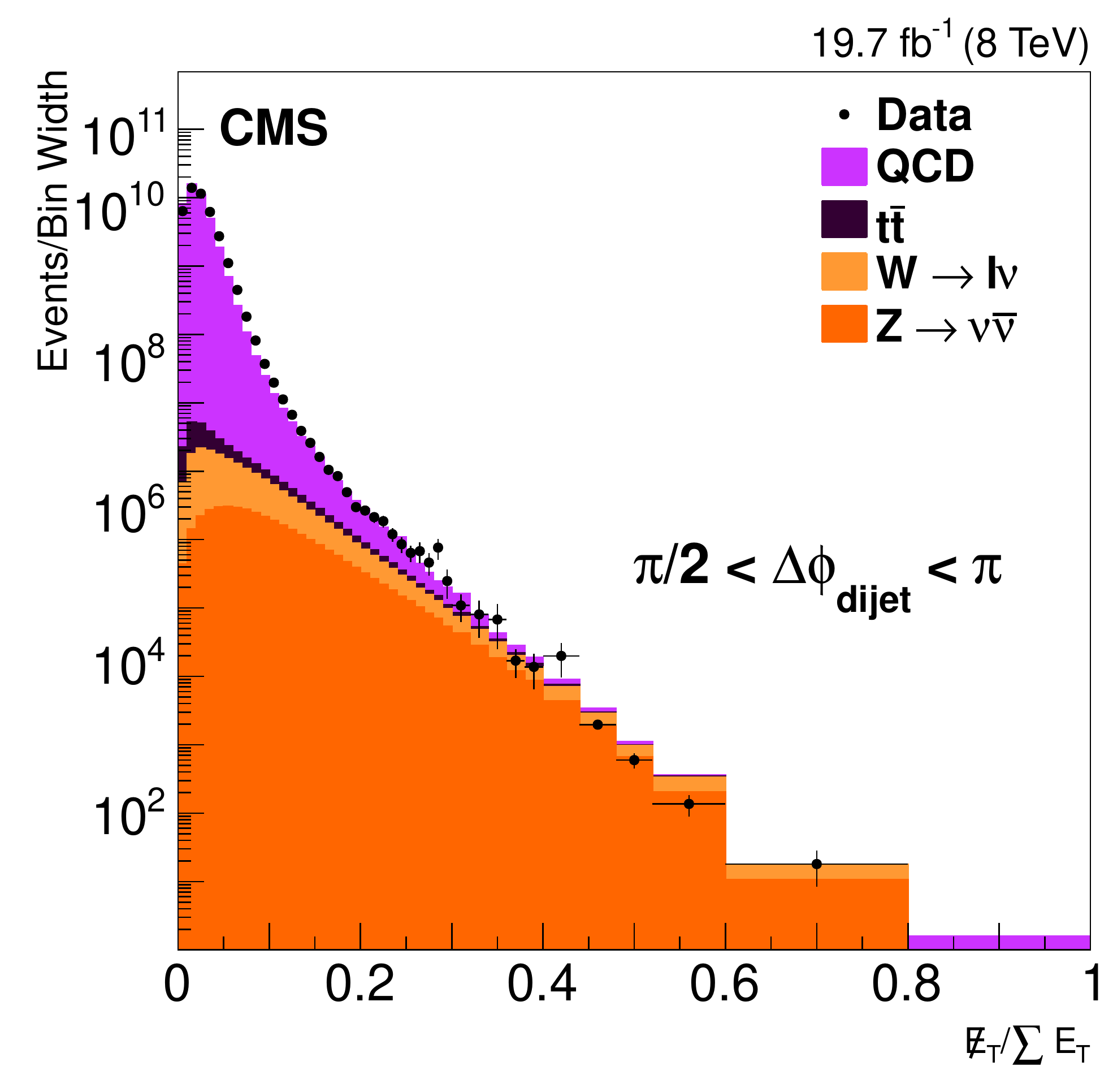}
  \caption{Distribution of \metoset for data (points) in comparison with
           simulated jet production and other processes with large \met
           (stacked) separately for the two regions $\dphi < \pi/2$
           (\cmsLeft) and $\pi/2 < \dphi < \pi$ (\cmsRight). The main contribution of
            events with large \met in the final state is caused by processes
            such as Z/W + jet(s) with $\Z \to \PGn \PAGn$ and
            $\PW\to \ell\PGn$.}
  \label{fig:stack_metoset}
\end{figure}

{\sloppy
To reduce the background from \ttbar and heavy vector boson
production, the variable \metoset is used.
The sum of the transverse energies is $\sum{\ET} = \sum_{i}E_{i}\sin\theta_{i}$,
and the missing transverse energy
$\met = \sqrt{ \left[\sum_{i}\left(E_{i}\sin\theta_{i}\cos\phi_{i}\right)\right]^2
+ \left[\sum_{i}\left(E_{i}\sin\theta_{i}\sin\phi_{i}\right)\right]^2 }$,
where $\theta$ is the polar angle and the sum runs over all PF candidates in the event.
A noticeable fraction of high-\pt jet events with large \met emerges from
\ttbar production with semileptonically decaying
b quarks. In addition, Z/W+jet(s) events with Z decays to neutrinos
and W decays into charged leptons with neutrinos have high \met values.
The distributions of the variable \metoset are shown in Fig.~\ref{fig:stack_metoset}
for the two regions $\dphi < \pi/2$ (\cmsLeft) and $\pi/2 < \dphi < \pi$ (\cmsRight).
The data (points) are compared to simulated events (stacked), using
\MADGRAPH 5.1.3.30 \cite{bib:madgraph5} matched to \PYTHIAS \cite{bib:pythia6}
for event generation.
Although some deviations of the simulation with respect to the data
are visible in Fig.~\ref{fig:stack_metoset} (cf.\
Ref. \cite{bib:JME_13_003}), the distributions allow a selection
criterion to be optimized with respect to the ratio of signal over
background. Events with $\metoset>0.1$ are rejected in both regions of
\dphi considered in Fig.~\ref{fig:stack_metoset}, which corresponds to
about 0.7\% of the data sample. Negligible background fractions of
$\approx$1\% and $\approx$0.1\% remain for the two regions $\dphi <
\pi/2$ and $\pi/2 < \dphi < \pi$, respectively.
\par}

\section{Measurement of the dijet cross section differential in
  \texorpdfstring{\dphi}{azimuthal angular difference}}
\label{sec:measurement}

The normalized dijet cross section differential in \dphi (Eq.~\ref{eq:dijet}) is corrected for detector
smearing effects and unfolded to the level of stable (decay length
$c\tau > 1\unit{cm}$) final-state particles. In this way, a direct comparison
of the measurement with corresponding results from other experiments and with
QCD predictions can be made.

The unfolding method is based on the matrix inversion algorithm implemented in the
software package \RooUnfold \cite{bib:RooUnfold}.
Unfolding uses a response matrix that maps the distribution at particle-level onto the measured one.
The response matrix is derived from a simulation that uses the true dijet cross
section distribution from \PYTHIAS with tune Z2* \cite{bib:Z2starCMS} as input, and introduces the
smearing effects by taking into account the \dphi resolution.
As a cross-check, the response matrix was filled from event samples
that have been passed through a detector simulation. No significant
difference was observed.
The unfolded distributions differ from the raw distributions
by 3 to 4\% for $\dphi < \pi/2$ and by less
than 3\% for $\pi/2 < \dphi < \pi$.
A two-dimensional unfolding based on the iterative D'Agostini algorithm \cite{bib:BayesUnfold},
which corrects for the smearing effects by taking into account both \dphi and \pt resolutions,
gives almost identical results.

The main systematic uncertainties arise from the estimation of the jet energy scale (JES)
calibration, the jet \pt resolution, and the unfolding correction.
The JES uncertainty is estimated to be 1.0--2.5\% for PF jets,
depending on the jet \pt and $\eta$ \cite{bib:JES_2012, bib:CMSasPDFs, bib:cms-jes}.
The resulting uncertainties in the normalized \dphi distributions range
from 7\% at $\dphi\approx 0$ via 3\% at $\pi/2$ to 1\% at $\pi$.

The jet \pt resolution is determined from a full detector simulation using events generated
by \PYTHIAS with tune Z2*, and is scaled by factors derived from data \cite{bib:cms-jes}.
The effect of the jet \pt resolution uncertainty is estimated by
varying it by one standard deviation up and down,
and comparing the \dphi distributions before and after the changes.
This results in a variation in the normalized \dphi distributions
ranging from 5\% at $\dphi\approx 0$ via 3\% at $\pi/2$ to 0.5\% at $\pi$.

The uncertainty in the unfolding correction factors is estimated by
checking the dependence of the response matrix on the choice of the Monte Carlo (MC)
generator. An alternative response matrix is built using the
\HERWIGpp 2.5.0 \cite{bib:herwigpp} event generator with the default tune of version 2.3.
The observed effect is less than 1\%. An additional systematic uncertainty obtained
by varying the \dphi resolution by $\pm$10\% to determine the unfolding
correction factors is estimated to be of the order of 1\%.
This variation of the \dphi resolution by $\pm$10\%
is motivated by the observed difference between data and
simulation in the \dphi resolution.
A total systematic unfolding uncertainty of 1\% accounts for the choice of
the MC generator in building the response matrix and the \dphi resolution.

\begin{figure}[hbtp]
  \centering
  \includegraphics[width=\cmsFigWidth]{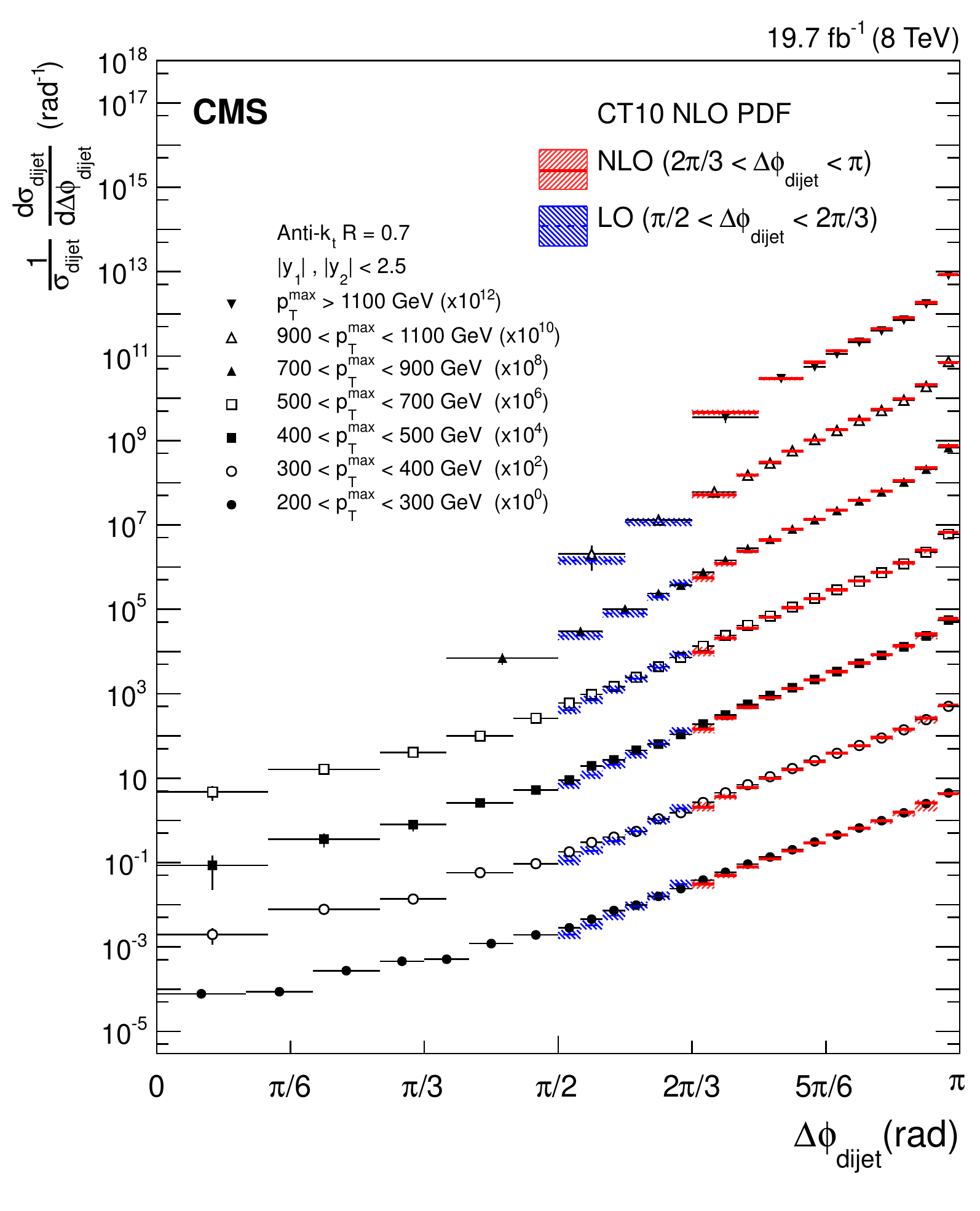}
  \caption{Normalized dijet cross section differential in \dphi for
    seven \ptmax regions, scaled by multiplicative factors for
    presentation purposes.  The error bars on the data points include
    statistical and systematic uncertainties. Overlaid on the data
    (points) are predictions from LO (dashed line; $\pi/2 \leq
    \dphi < 2\pi/3$) and NLO (solid line; $2\pi/3 \leq \dphi \leq
    \pi$) calculations using the CT10 NLO PDF set.
    The PDF, \alps, and scale uncertainties are
    added in quadrature to give the total theoretical uncertainty, which
    is indicated by the downwards-diagonally (LO) and
    upwards-diagonally (NLO) hatched regions around the theory lines.}
  \label{fig:xsection_particle_level}
\end{figure}

The unfolded dijet cross section differential in \dphi and normalized
by the dijet cross section integrated over the entire phase space is
shown in Fig.~\ref{fig:xsection_particle_level} for seven \ptmax
regions. Each region is scaled by a multiplicative factor for
presentation purposes. The \dphi distributions are strongly peaked at
$\pi$ and become steeper with increasing \ptmax.
Overlaid on the data for $\dphi > \pi/2$ are predictions from pQCD,
presented in more detail in the next section, using parton distribution
functions (PDF) of the CT10 PDF set.

\section{Comparison to theoretical predictions}
\label{sec:theory}
\subsection{Predictions from fixed-order calculations in pQCD}
\label{sec:fixedorder}

The theoretical predictions for the normalized dijet cross section
differential in \dphi are based on a 3-jet calculation at NLO\@. The
correction of nonperturbative (NP) effects, which account for multiparton
interactions (MPI) and hadronization,
is studied using event
samples simulated with the \PYTHIAS (tune Z2*) and \HERWIGpp
(tune~2.3) event generators. Small NP effects are expected,
since this measurement deals with a normalized distribution. These
corrections are found to be of the order of 1\%, roughly at the limit
of the accuracy of the MC simulations. Therefore NP corrections are
considered to be negligible and are not applied.

The fixed-order calculations are performed using the \NLOJETPP program
version~4.1.3 \cite{Nagy:2001fj,Nagy:2003tz} within the framework of
the \fastNLO package version~2.3.1 \cite{Britzger:2012bs}. The
differential cross section is calculated for 3-jet production at NLO,
\ie up to terms of order $\alps^{4}$, with three or four partons in
the final state. This calculation has LO precision in the region
$\pi/2 \leq \dphi < 2\pi/3$ and NLO precision for $2\pi/3 \leq \dphi <
\pi$.
The bin including $\dphi = \pi$ is computed from the NLO dijet cross
section within this bin. %
For each region in \ptmax, the differential cross section is normalized
to the dijet cross section calculated at LO for $\pi/2 \leq \dphi <
2\pi/3$ and at NLO, \ie up to terms proportional to $\alps^{3}$, for
$2\pi/3 \leq \dphi \leq \pi$.
The use of the LO dijet cross section for
the normalization in the region $\pi/2 \leq \dphi < 2\pi/3$ leads to
an improved description of the data and avoids artificially increased
scale uncertainties as described in
Refs. \cite{Wobisch:2015jea,Seymour:1997kj}. Of course, this
difference in normalization leads to a discontinuity proportional to
$\sigma_\text{dijet}^\mathrm{NLO}/\sigma_\text{dijet}^\mathrm{LO}$
at $\dphi = 2\pi/3$.

The number of quark flavours that are assumed to be massless
is set to five, and the renormalization and factorization scales, \mur and
\muf, are chosen to be equal to \ptmax. The PDF sets with NLO
evolutions used in the calculations are tabulated in
Table~\ref{tab:pdfsets}. The ABM11 PDF set utilizes a fixed flavour
number scheme, whereas the rest of the PDF sets use a variable flavour
number scheme. The maximum number of flavours is denoted by
$N_f$.

\begin{table}[htbp]
  \topcaption{The PDF sets used to compare the data with expectations, together with the corresponding
    maximum number of flavours $N_f$
    and the default values of \alpsmz.}
  \label{tab:pdfsets}
  \centering
  \begin{tabular}{llrr}
    \hline
    Base set   & Refs. \ & $N_f$  &
    \alpsmz \\
    \hline
    ABM11      & \cite{Alekhin:2012ig}               &       5       & 0.1180 \\
    CT10       & \cite{Lai:2010vv}                   & ${\leq}$5    & 0.1180 \\
    HERAPDF1.5 & \cite{Aaron:2009aa}                 & ${\leq}$5    & 0.1176 \\
    MSTW2008   & \cite{Martin:2009iq}                & ${\leq}$5    & 0.1202 \\
    NNPDF21    & \cite{Ball:2011mu}                  & ${\leq}$6    & 0.1190 \\
    \hline
  \end{tabular}
\end{table}

{\tolerance=800
The uncertainties due to the renormalization and factorization scales
are evaluated by varying the default choice of $\mur = \muf = \ptmax$
between \ptmax/2 and 2\ptmax, simultaneously in the differential cross
section and in the total cross section, in the following six
combinations: $(\mur/\ptmax , \muf/\ptmax) = (1/2,1/2)$, $(1/2,1)$,
$(1,1/2)$, $(1,2)$, $(2,1)$, and $(2,2)$.  The PDF uncertainties are evaluated
according to the prescriptions for the CT10 PDF set in
Ref. \cite{Pumplin:2002vw}. The CT10 PDF set employs the eigenvector
method with upward and downward variations for each eigenvector. To
evaluate the uncertainty due to the value of the strong coupling
constant at 68\% confidence level, \alpsmz is varied by $\pm$0.001 as recommended in Ref. \cite{Lai:2010nw}.
\par}

\begin{figure*}[hbtp]
  \centering
  \includegraphics[width=0.75\textwidth]{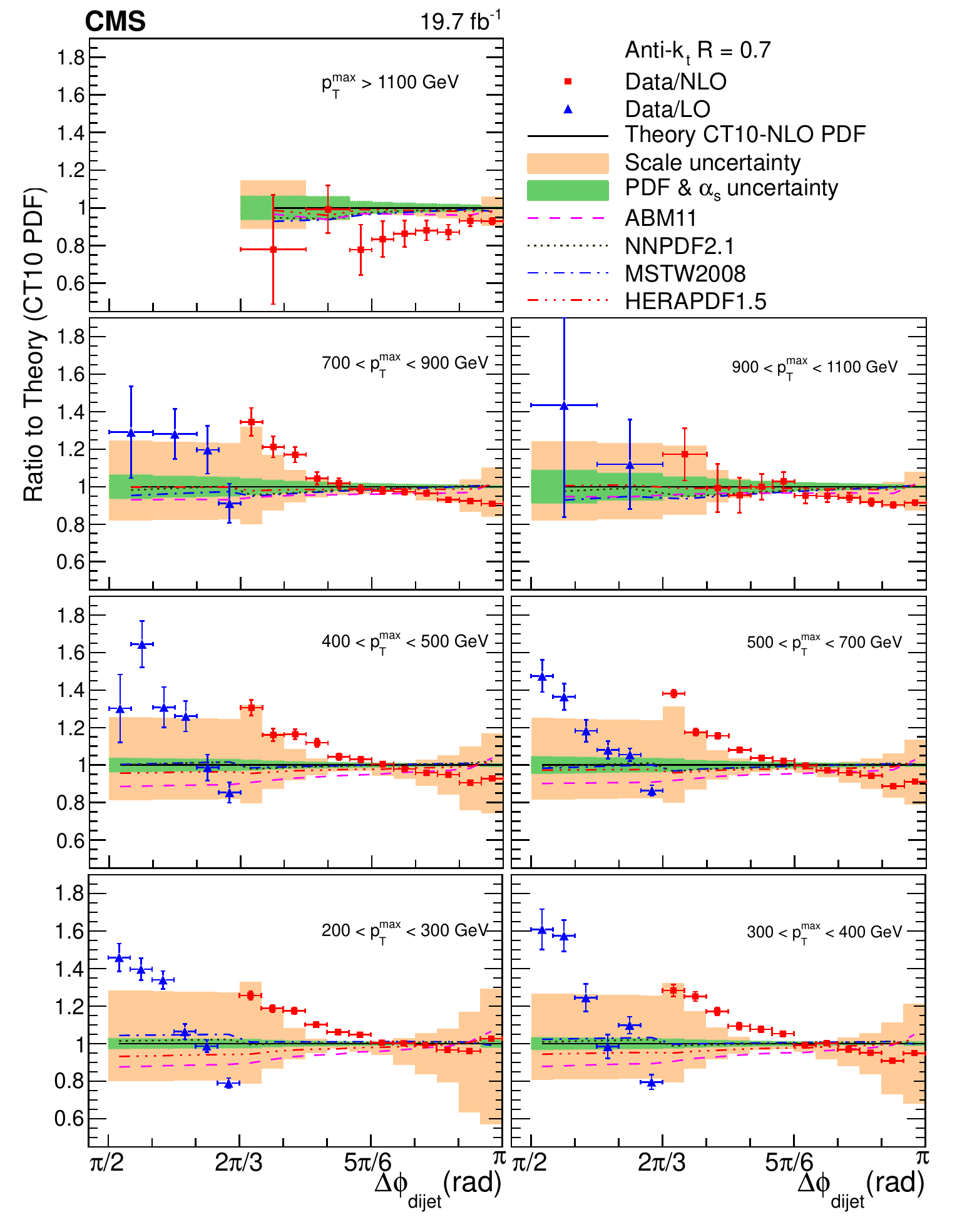}
  \caption{Ratios of the normalized dijet cross section differential
    in \dphi to LO (triangles) and NLO (squares) pQCD
    predictions using the CT10 PDF set at next-to-leading evolution
    order for all \ptmax regions. The error bars on the data points
    represent the total experimental uncertainty, which is the
    quadratic sum of the statistical and systematic uncertainties. The
    uncertainties of the theoretical predictions are shown as
    inner band (PDF \& \alps) and outer band (scales). The
    predictions using various other PDF sets relative to CT10 are
    indicated with different line styles.}
  \label{fig:DataTheory_CT10nlo}
\end{figure*}

The results of fixed-order calculations with the CT10 PDF set are
overlaid on the data for $\dphi > \pi/2$ in
Fig.~\ref{fig:xsection_particle_level}.
Figure \ref{fig:DataTheory_CT10nlo} shows the ratio
of the normalized dijet cross section differential in \dphi to
theory calculated using the CT10 PDF set,
together with the combined PDF and \alps uncertainty (inner band), and
the scale uncertainty (outer band).
Also shown are the ratios of theory derived with the alternative PDF
sets ABM11 (dashed line), HERAPDF1.5 (dashed--three-dotted line),
MSTW2008 (dashed-dotted line), and NNPDF2.1 (dotted line) compared to
the prediction with the CT10 PDFs.

The fixed-order calculations agree with the data for azimuthal angular
separations larger than $5\pi/6$ except for the highest \ptmax region,
where they exceed the data. For smaller \dphi values between $2\pi/3$
and $5\pi/6$, in particular where the estimate of the theoretical
uncertainties becomes small, systematic discrepancies are exhibited
that diminish with increasing \ptmax. In the 4-jet LO region with
$\dphi < 2\pi/3$, the pattern of increasing deviations towards smaller
\dphi and decreasing deviations towards larger \ptmax is repeated, but
with less significance because of the larger scale uncertainty.
Similar observations were made in the previous CMS
measurement \cite{bib:CMS}, which exhibited larger discrepancies in the
4-jet region due to the normalization to the NLO dijet
cross section instead of a LO one.

\subsection{Predictions from fixed-order calculations matched to parton shower simulations}
\label{sec:compMC}

The \PYTHIAS \cite{bib:pythia6}, \PYTHIAE \cite{bib:pythia8}, and
\HERWIGpp \cite{bib:herwigpp} event generators complement LO
dijet matrix elements with parton showers to simulate higher-order
processes.
Both \PYTHIA versions, \PYTHIAS with the Z2* tune \cite{bib:Z2starCMS}
and \PYTHIAE with the CUETM1 tune \cite{bib:GEN-14-001},
employ \pt-ordered parton showers \cite{Sjostrand:2004ef,bib:pythia8tune4c},
while \HERWIGpp with the default tune of version~2.3 uses a coherent-branching
algorithm with angular ordering of the showers \cite{Gieseke:2003rz}.

The \MADGRAPH program version~5.1.5.7 \cite{bib:madgraph5} supplies the results of LO
matrix element calculations with two to four outgoing partons that can
be matched to the implementations of parton showers, hadronization, and
MPI of the event generators. In this
analysis, it is interfaced with \PYTHIAS with tune Z2* using the MLM
matching procedure \cite{bib:mlm}
to avoid any double counting between tree-level and
parton shower generated parton configurations.

The \POWHEG framework
\cite{bib:Frixione:2007vw,bib:Alioli:2010xd,bib:Nason:2004rx} provides
an NLO dijet calculation \cite{bib:POWHEG_Dijet} that can also be
matched via the parton showers to event generators. Here, \POWHEG is
used with the CT10NLO PDF set and is interfaced to \PYTHIAE with the
CUET \cite{bib:GEN-14-001} tune, which employs the LO CTEQ6L1
\cite{Pumplin:2002vw} PDF set.
Predictions with parton showers matched to a NLO 3-jet calculation
using \POWHEG \cite{Kardos:2014dua} or \MGaMCNLO \cite{Alwall:2014hca}
would be even more relevant for a multijet topology.
They could not, however, be included within the timescale of this
analysis. %
Approaching azimuthal angular separations close to $\pi$, it might
also be interesting to compare to predictions employing the technique
of \pt resummation~\cite{Sun:2015doa}.

\begin{figure}[hbtp]
  \centering
  \includegraphics[width=\cmsFigWidth]{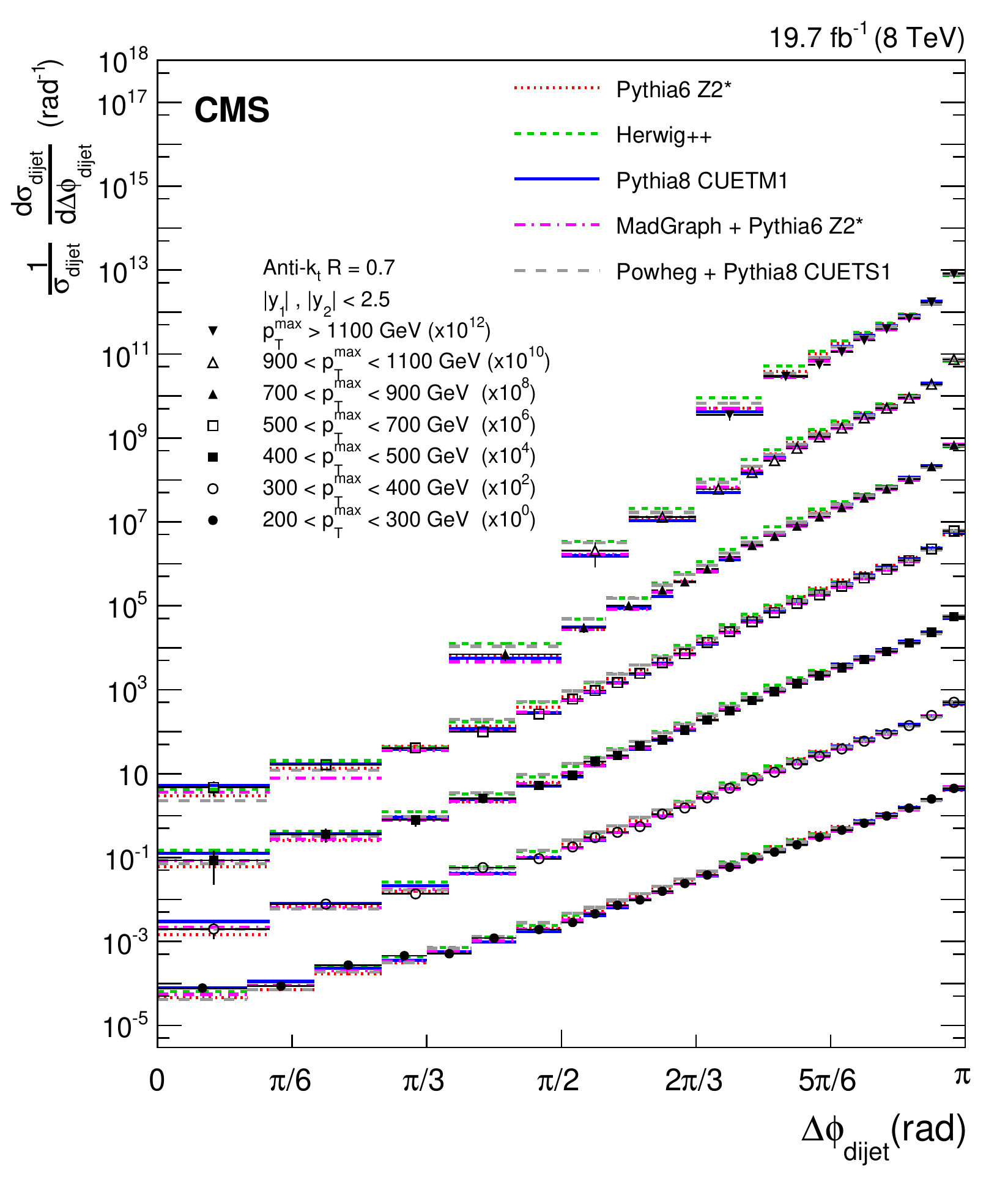}
  \caption{Normalized dijet cross section differential in \dphi
    for seven \ptmax regions, scaled by multiplicative factors for
    presentation purposes.  The error bars on the data points include
    statistical and systematic uncertainties.  Overlaid on the data
    are predictions from the \PYTHIAS, \HERWIGpp, \PYTHIAE,
    \MADGRAPH + \PYTHIAS, and \POWHEG + \PYTHIAE event generators.}
  \label{fig:xsection_MC_data}
\end{figure}

\begin{figure*}[hbtp]
  \centering
  \includegraphics[width=0.75\textwidth]{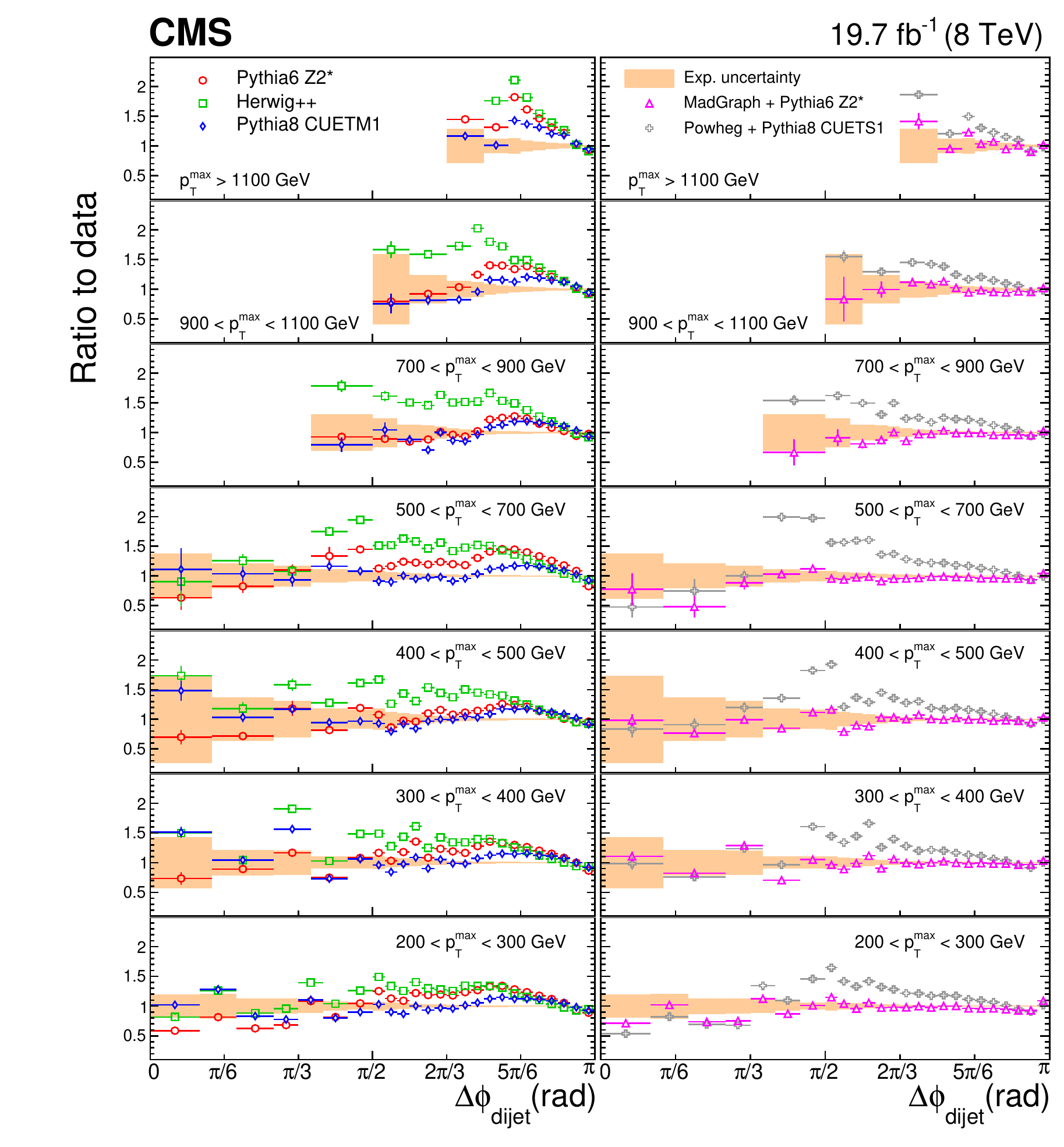}
  \caption{Ratios of \PYTHIAS, \HERWIGpp, \PYTHIAE, \MADGRAPH + \PYTHIAS, and
    \POWHEG + \PYTHIAE predictions to the normalized dijet cross
    section differential in \dphi, for all \ptmax regions.
    The solid band indicates the total experimental uncertainty and
    the error bars on the MC points represent the statistical
    uncertainties of the simulated data.}
  \label{fig:ratios_MC_data}
\end{figure*}

In Fig.~\ref{fig:xsection_MC_data} the normalized dijet cross section
differential in \dphi is compared to the predictions from fixed-order
calculations supplemented with parton showers, hadronization, and MPI\@.
The error bars on the data points represent the total experimental uncertainty,
which is the quadratic sum of the statistical and systematic uncertainties.
Figure~\ref{fig:ratios_MC_data} shows the ratios of these predictions to
the normalized dijet cross section differential in \dphi, for the seven \ptmax regions.
The solid band indicates the total experimental uncertainty and the error
bars on the MC points represent the statistical uncertainties in the simulated data.

{\sloppy
Among the LO dijet event generators \PYTHIAS, \PYTHIAE, and
\HERWIGpp, \PYTHIAE exhibits the smallest deviations from the measurements.
\PYTHIAS and \HERWIGpp systematically overshoot the data, particular around
$\dphi = 5\pi/6$. The best description of the measurement is
given by the tree-level multiparton event generator \MADGRAPH
interfaced with \PYTHIAS for showering, hadronization, and
MPI\@.
The \POWHEG generator (here used only in the NLO dijet mode) matched to \PYTHIAE
shows deviations from the data similar to the LO dijet event generators.\par}

\section{Summary}
\label{sec:summary}

A measurement is presented of the normalized dijet cross section differential in
the azimuthal angular separation \dphi of the two jets leading in \pt
for seven regions in the leading-jet transverse momentum
\ptmax. The data set of pp collisions at 8\TeV centre-of-mass energy
collected in 2012 by the CMS experiment and corresponding to an integrated luminosity of
19.7\fbinv is analysed.

The measured distributions in \dphi are compared to calculations in
perturbative QCD for 3-jet production with up to four outgoing partons
that provide NLO predictions for the range of
$2\pi/3 \leq \dphi < \pi$ and LO predictions for $\pi/2 \leq \dphi <
2\pi/3$. The NLO predictions describe the data down to values of
$\dphi\approx 5\pi/6$, but deviate increasingly when approaching the
4-jet region, starting at $\dphi = 2\pi/3$, particularly at low
\ptmax. The pattern of increasing deviations towards smaller \dphi and
decreasing deviations towards larger \ptmax is repeated in the 4-jet
LO region with $\dphi < 2\pi/3$, but with less significance because of
the larger scale uncertainty.

In a comparison of the normalized \dphi distributions to the LO
dijet event generators \PYTHIAS, \PYTHIAE, and \HERWIGpp, \PYTHIAE
gives the best agreement. \PYTHIAS and \HERWIGpp systematically
overshoot the data, particularly for $\dphi \approx 5\pi/6$.
A good overall description of the measurement is provided by the tree-level
multijet event generator \MADGRAPH in combination with \PYTHIAS for
showering, hadronization, and multiparton interactions.
The dijet NLO calculations from \POWHEG matched to \PYTHIAE exhibit
deviations similar to the LO dijet event generators. Improved
multijet predictions can be expected from 3-jet NLO calculations
matched to parton showers like from \POWHEG or \MGaMCNLO.

Similar observations were reported previously by CMS \cite{bib:CMS}
and ATLAS \cite{bib:ATLAS}, but with less significance because of the
smaller data sets. The extension to \dphi values below $\pi/2$,
the improved LO description in the 4-jet region $\pi/2 \leq \dphi <
2\pi/3$, and the comparison to dijet NLO calculations matched to parton showers
are new results of the present analysis.

\begin{acknowledgments}
We acknowledge discussions and comparisons with P.~Sun, C.~P.~Yuan, and F.~Yuan following the approach of~\cite{Sun:2015doa}.

We congratulate our colleagues in the CERN accelerator departments for the excellent performance of the LHC and thank the technical and administrative staffs at CERN and at other CMS institutes for their contributions to the success of the CMS effort. In addition, we gratefully acknowledge the computing centres and personnel of the Worldwide LHC Computing Grid for delivering so effectively the computing infrastructure essential to our analyses. Finally, we acknowledge the enduring support for the construction and operation of the LHC and the CMS detector provided by the following funding agencies: BMWFW and FWF (Austria); FNRS and FWO (Belgium); CNPq, CAPES, FAPERJ, and FAPESP (Brazil); MES (Bulgaria); CERN; CAS, MoST, and NSFC (China); COLCIENCIAS (Colombia); MSES and CSF (Croatia); RPF (Cyprus); MoER, ERC IUT and ERDF (Estonia); Academy of Finland, MEC, and HIP (Finland); CEA and CNRS/IN2P3 (France); BMBF, DFG, and HGF (Germany); GSRT (Greece); OTKA and NIH (Hungary); DAE and DST (India); IPM (Iran); SFI (Ireland); INFN (Italy); MSIP and NRF (Republic of Korea); LAS (Lithuania); MOE and UM (Malaysia); CINVESTAV, CONACYT, SEP, and UASLP-FAI (Mexico); MBIE (New Zealand); PAEC (Pakistan); MSHE and NSC (Poland); FCT (Portugal); JINR (Dubna); MON, RosAtom, RAS and RFBR (Russia); MESTD (Serbia); SEIDI and CPAN (Spain); Swiss Funding Agencies (Switzerland); MST (Taipei); ThEPCenter, IPST, STAR and NSTDA (Thailand); TUBITAK and TAEK (Turkey); NASU and SFFR (Ukraine); STFC (United Kingdom); DOE and NSF (USA).

Individuals have received support from the Marie-Curie programme and the European Research Council and EPLANET (European Union); the Leventis Foundation; the A. P. Sloan Foundation; the Alexander von Humboldt Foundation; the Belgian Federal Science Policy Office; the Fonds pour la Formation \`a la Recherche dans l'Industrie et dans l'Agriculture (FRIA-Belgium); the Agentschap voor Innovatie door Wetenschap en Technologie (IWT-Belgium); the Ministry of Education, Youth and Sports (MEYS) of the Czech Republic; the Council of Science and Industrial Research, India; the HOMING PLUS programme of the Foundation for Polish Science, cofinanced from European Union, Regional Development Fund; the OPUS programme of the National Science Center (Poland); the Compagnia di San Paolo (Torino); MIUR project 20108T4XTM (Italy); the Thalis and Aristeia programmes cofinanced by EU-ESF and the Greek NSRF; the National Priorities Research Program by Qatar National Research Fund; the Rachadapisek Sompot Fund for Postdoctoral Fellowship, Chulalongkorn University (Thailand); the Chulalongkorn Academic into Its 2nd Century Project Advancement Project (Thailand); and the Welch Foundation, contract C-1845.
\end{acknowledgments}
\clearpage
\bibliography{auto_generated}
\cleardoublepage \appendix\section{The CMS Collaboration \label{app:collab}}\begin{sloppypar}\hyphenpenalty=5000\widowpenalty=500\clubpenalty=5000\textbf{Yerevan Physics Institute,  Yerevan,  Armenia}\\*[0pt]
V.~Khachatryan, A.M.~Sirunyan, A.~Tumasyan
\vskip\cmsinstskip
\textbf{Institut f\"{u}r Hochenergiephysik der OeAW,  Wien,  Austria}\\*[0pt]
W.~Adam, E.~Asilar, T.~Bergauer, J.~Brandstetter, E.~Brondolin, M.~Dragicevic, J.~Er\"{o}, M.~Flechl, M.~Friedl, R.~Fr\"{u}hwirth\cmsAuthorMark{1}, V.M.~Ghete, C.~Hartl, N.~H\"{o}rmann, J.~Hrubec, M.~Jeitler\cmsAuthorMark{1}, V.~Kn\"{u}nz, A.~K\"{o}nig, M.~Krammer\cmsAuthorMark{1}, I.~Kr\"{a}tschmer, D.~Liko, T.~Matsushita, I.~Mikulec, D.~Rabady\cmsAuthorMark{2}, N.~Rad, B.~Rahbaran, H.~Rohringer, J.~Schieck\cmsAuthorMark{1}, R.~Sch\"{o}fbeck, J.~Strauss, W.~Treberer-Treberspurg, W.~Waltenberger, C.-E.~Wulz\cmsAuthorMark{1}
% ...
\vskip\cmsinstskip
% ...
1:~~Also at Vienna University of Technology, Vienna, Austria\\
2:~~Also at CERN, European Organization for Nuclear Research, Geneva, Switzerland\\
% ...

\end{sloppypar}
\end{document}